\documentclass[prb,superscriptaddress, floatfix, letterpaper, twocolumn, aps, showpacs]{revtex4-1}
\usepackage[utf8]{inputenc}
\usepackage{natbib}
\usepackage{amsmath}
\usepackage{amsbsy}
\usepackage{amssymb}
\usepackage{scrextend}
\usepackage{graphicx}
\usepackage{color}
\usepackage{xcolor}
\usepackage{float}

\newcommand{\RbEu}{$\mathrm{Rb}\mathrm{Eu}\mathrm{Fe}_{4}\mathrm{As}_{4}$}
\newcommand{\RbEuNix}{$\mathrm{Rb}\mathrm{Eu}(\mathrm{Fe}_{1-x}\mathrm{Ni}_{x})_{4}\mathrm{As}_{4}$}

\begin{document}
\title{Magnetic and superconducting anisotropy in Ni-doped RbEuFe$_4$As$_4$ single crystals}
\author{K. Willa}
\affiliation{Materials Science Division, Argonne National Laboratory, 9700 South Cass Avenue, Lemont, IL 60439, USA}
\affiliation{Institute for Solid State Physics, Karlsruhe Institute of Technology, Karlsruhe D-76021, Germany }
\author{M.P. Smylie}
\affiliation{Materials Science Division, Argonne National Laboratory, 9700 South Cass Avenue, Lemont, IL 60439, USA}
\affiliation{Department of Physics and Astronomy, Hofstra University, Hempstead, New York 11549}
\author{Y. Simsek}
\affiliation{Materials Science Division, Argonne National Laboratory, 9700 South Cass Avenue, Lemont, IL 60439, USA}
\affiliation{SUNUM-Sabanci University Nanotechnology Research and Application Center, Istanbul, Turkey}
\author{J.-K.\ Bao}
\affiliation{Materials Science Division, Argonne National Laboratory, 9700 South Cass Avenue, Lemont, IL 60439, USA}
\author{D. Y.\ Chung}
\affiliation{Materials Science Division, Argonne National Laboratory, 9700 South Cass Avenue, Lemont, IL 60439, USA}
\author{M. G.\ Kanatzidis}
\affiliation{Materials Science Division, Argonne National Laboratory, 9700 South Cass Avenue, Lemont, IL 60439, USA}
\affiliation{Department of Chemistry, Northwestern University, Evanston, Illinois, 60208, USA}
\author{W.-K.\ Kwok}
\affiliation{Materials Science Division, Argonne National Laboratory, 9700 South Cass Avenue, Lemont, IL 60439, USA}
\author{U.\ Welp}
\affiliation{Materials Science Division, Argonne National Laboratory, 9700 South Cass Avenue, Lemont, IL 60439, USA}

\date{\today}

\begin{abstract}
We investigate the effect of Ni doping on the Fe-site in single crystals of the magnetic superconductor {\RbEu} for doping concentrations of up to 4\%. A clear suppression in the superconducting transition temperature is observed in specific heat, resistivity and magnetization measurements. Upon Ni-doping, the resistivity curves shift up in a parallel fashion indicating a strong increase of the residual resistivity due to scattering by charged dopand atoms while the shape of the curve and thus the electronic structure appears largely unchanged. The observed step $\Delta C/T_c$ at the superconducting transition decreases strongly for increasing Ni doping in agreement with expectations based on a model of multi-band superconductivity and strong inter-band pairing. The upper critical field slopes are reduced upon Ni doping for in- as well as out-of-plane fields leading to a small reduction in the superconducting anisotropy. The specific heat measurements of the magnetic transition reveal the same BKT behavior close to the transition temperature $T_m$ for all doping levels. The transition temperature is essentially unchanged upon doping. The in to out-of-plane anisotropy of Eu-magnetism observed at small magnetic fields is unaltered as compared to the undoped compound. All of these observations indicate a decoupling of the Eu magnetism from superconductivity and essentially no influence of Ni doping on the Eu magnetism in this compound.
\end{abstract}

\maketitle
\section{Introduction}

The family of iron-based superconductors has been constantly growing, now comprising many members of the so-called 11, 111, 1111 and 122 materials \cite{Kamihara2008, Rotter2008, Lynn2009, Johnston2010, Stewart2012, Tanabe2011, Hosono2015, Si2016, Bang2017}. In the majority of these materials, superconductivity arises from an antiferromagnetic parent compound upon electron, hole or isovalent doping or mechanical pressure \cite{Kamihara2008, Rotter2008, Johnston2010, Stewart2012, Tanabe2011, Hosono2015, Si2016, Bang2017, Kurita2011, Hardy2010, Meingast2012, Hardy2016, Jin2019}. The so-called 1144 compounds with composition $A Ae \mathrm{Fe}_4\mathrm{As}_4$ ($A = \mathrm{K}$, Rb, Cs; $Ae = \mathrm{Ca}$, Sr, Eu) \cite{Iyo2016, Kawashima2016, Liu2016a, Liu2016b, Meier2016} form the most recently discovered members of this family. In these materials, the large difference in the ionic radii of the A and Ae atoms leads to the formation of alternating A and Ae-layers between the Fe$_2$As$_2$-planes producing an asymmetric environment for the Fe$_2$As$_2$-layers. Contrary to the doped 122 compounds, the 1144 compounds are superconducting in their stoichiometric state with $T_c$ reaching $\approx$ 37 K. A formal charge count yields that stoichiometric 1144 are intrinsically hole doped to $\approx$ 0.25 holes/Fe, close to the doping level that yields optimum $T_c$ in the 122 compounds. Recently, it has been realized \cite{Liu2019} that the high purity and symmetry properties of 1144 materials such as CaKFe$_4$As$_4$ may open a new platform for the observation of topological bandstructures and Majorana zero modes. The $A$EuFe$_4$As$_4$ compounds represent a peculiar subgroup of this new family as the planes of Eu$^{2+}$ ions can order magnetically as has been shown previously in EuFe$_2$As$_2$ \cite{Jeevan2011, Koo2010, Zapf2017, Nandi2014, Xiao2009}. Initial experiments on polycrystalline \RbEu samples indicate ferromagnetic in-plane ordering at $15 \mathrm{K}$ deep within the superconducting phase ($T_c = 37$ K) \cite{Liu2016a, Liu2016b}. Recent studies on single crystals \cite{Bao2018} affirmed these findings and allowed for the study of the magnetic and superconducting anisotropy \cite{Smylie2018c, Willa2019}, demonstrating a low superconducting anisotropy of 1.8, highly anisotropic quasi-2D Eu-magnetism, and an associated Berezinskii-Kosterlitz-Thouless (BKT) transition of the Eu$^{2+}$ moments. Large upper critical fields were observed in pulsed field measurements \cite{Smylie2019}. Recent neutron scattering data indicate a modulated, possibly helical stacking of ferromagnetic Eu-layers along the $c$ axis with a period of four $c$-axis lattice constants \cite{Iida2019}.
Furthermore, a study on polycrystalline samples \cite{Liu2017} revealed that $T_c$ is suppressed upon Ni-substitution on the Fe-site, while the magnetic ordering temperature is unchanged. Meanwhile, Ca-substitution on the Eu-site \cite{Kawashima2018} suppressed the magnetic ordering temperature in polycrystalline samples without changing $T_c$. These results and recent high pressure studies \cite{Jackson2018} suggest an almost complete decoupling of superconductivity hosted by the Fe$_2$As$_2$ planes and magnetism within the Eu layers. This model system therefore provides an ideal playground to selectively tune the relative importance of magnetism and superconductivity, rendering these compounds promising for further investigations. In this work we investigate single crystals of Ni-doped {\RbEu} with a particular focus on the anisotropy of the superconducting and magnetic transition and its doping dependence.

\section{Experimental Methods}
\begin{figure}[tbh]
\includegraphics[width=0.98\linewidth]{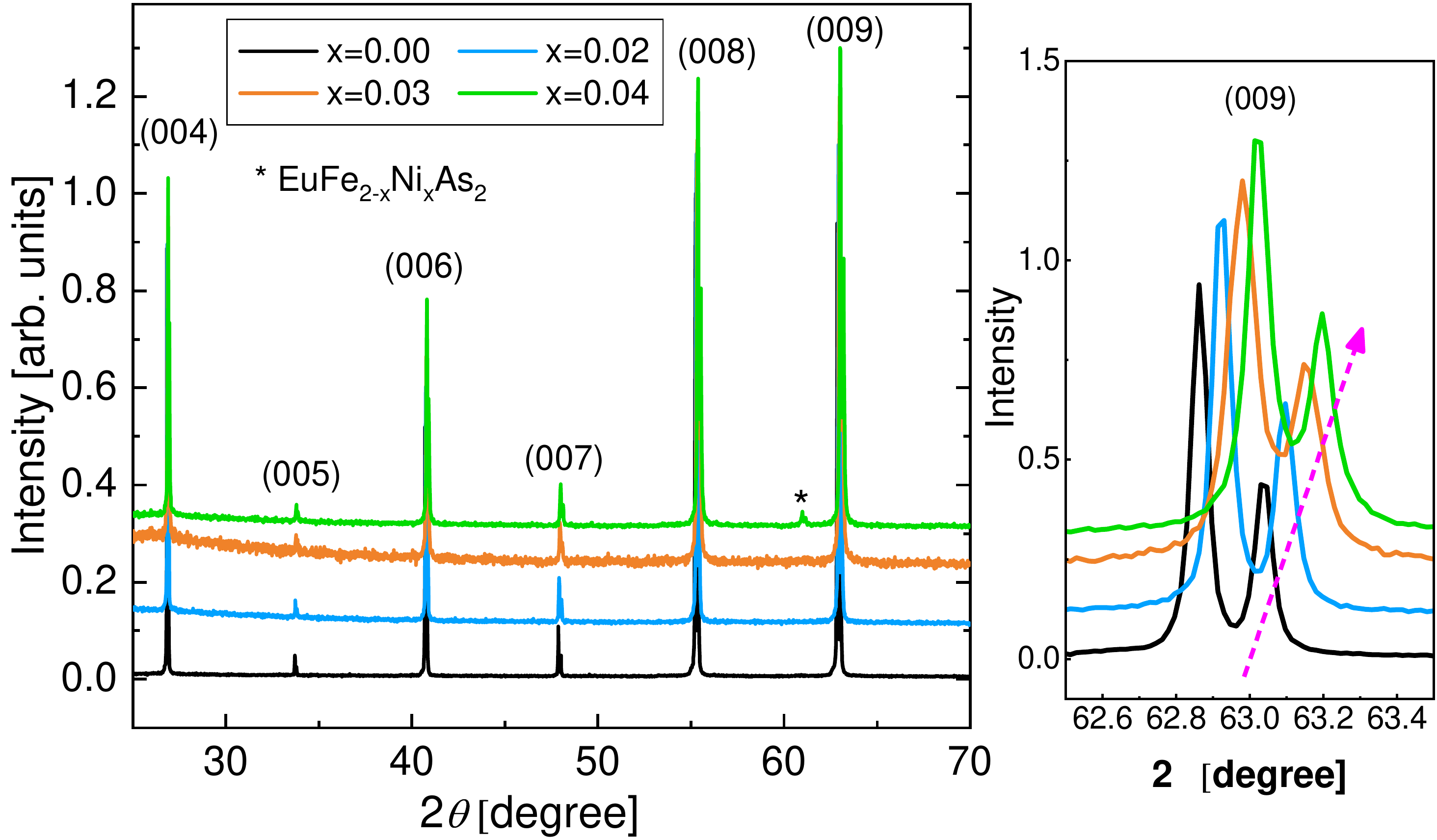}
\caption{XRD measurements of (00l) peaks on single-crystal {\RbEuNix} for $x = 0.00$ (black), $x = 0.02$ (blue), $x = 0.03$ (orange), $x = 0.04$ (green). As doping increases, the lattice peaks systematically shift to higher angles as the lattice shrinks; compare table \ref{tab:data}. The inset shows the (009) peak. The asterisk marks a small amount of 122-phase compound that grows at the surface of the 1144-phase. No other phases have been observed.
}
\label{fig:xray}
\end{figure}

High quality single crystals {\RbEuNix} were grown out of RbAs flux with a similar method to the one used for the undoped {\RbEu} in Ref. \cite{Bao2018}. In order to provide the Ni doping, an extra NiAs precursor was synthesized by reacting Ni and As powder under a 1:1 ratio in an evacuated quartz tube at $823 \mathrm{K}$ for 12 hours. The sintered NiAs precursor was then ground into a fine powder in an agate mortar. To achieve a nominal Ni doping $x$ ($x = 0.03$, 0.05, 0.07), EuAs:Fe$_2$As:NiAs:RbAs = 1:1-$x$:2$x$:15 were used for the RbAs flux growth. After the growth the RbEu(Fe$_{1-x}$Ni$_x$)$_4$As$_4$ crystals are obtained by removing the RbAs flux with reagent alcohol. This procedure yields thin flat plates with typical sizes of $300 \times 300 \times 50 \mu$m$^3$, with the tetragonal [110] and $[1\overline{1}0]$ orientations parallel to the long edges, and the [001] orientation perpendicular to the plate.

We performed x-ray diffraction (XRD) measurements on single crystals using a Powder X-ray PANalytical diffractometer X'pert with a Cu-K$_{\alpha}$ source. Plate-like crystals were well oriented in a flat zero-background single crystalline silicon sample holder to make sure that peaks with the index (00l) were observed, see Figure \ref{fig:xray}. Magnetization measurements were performed in a Quantum Design MPMS-7 system with samples mounted in a notched quartz rod for $H \| [001]$ and on a smooth quartz rod for $H \| [110]$. For low-field measurements, careful background field calibration was performed prior to each run with a Pd reference sample. For magnetotransport measurements, four-bar gold patterns were deposited onto thin rectangular-shaped crystals with the help of a shadow-mask technique then thin gold wires were attached onto the gold bars with silver epoxy. The measurements were performed inside a $9-1-1 \mathrm{T}$ 3-axis AMI superconducting vector magnet in a 4-point geometry and applying DC currents of 1 mA along the in-plane direction. The ac specific heat is measured on a SiN calorimetric membrane \cite{Tagliati2012, Willa2017}. For this purpose, a small ($\approx 100 \times 100 \times 25 \mu$m$^3$) platelet-shaped single crystal was mounted on the nanocalorimeter platform with apiezon grease and inserted into the same three axis superconducting vector magnet as used for resistivity measurements. The sample was then subjected to a small oscillatory heating current with frequencies usually in the range of 1Hz and the response of the sample was measured using a SynkTek MCL1-540 multichannel lock-in system.
\begin{figure}[t!]
\includegraphics[width=0.82\linewidth]{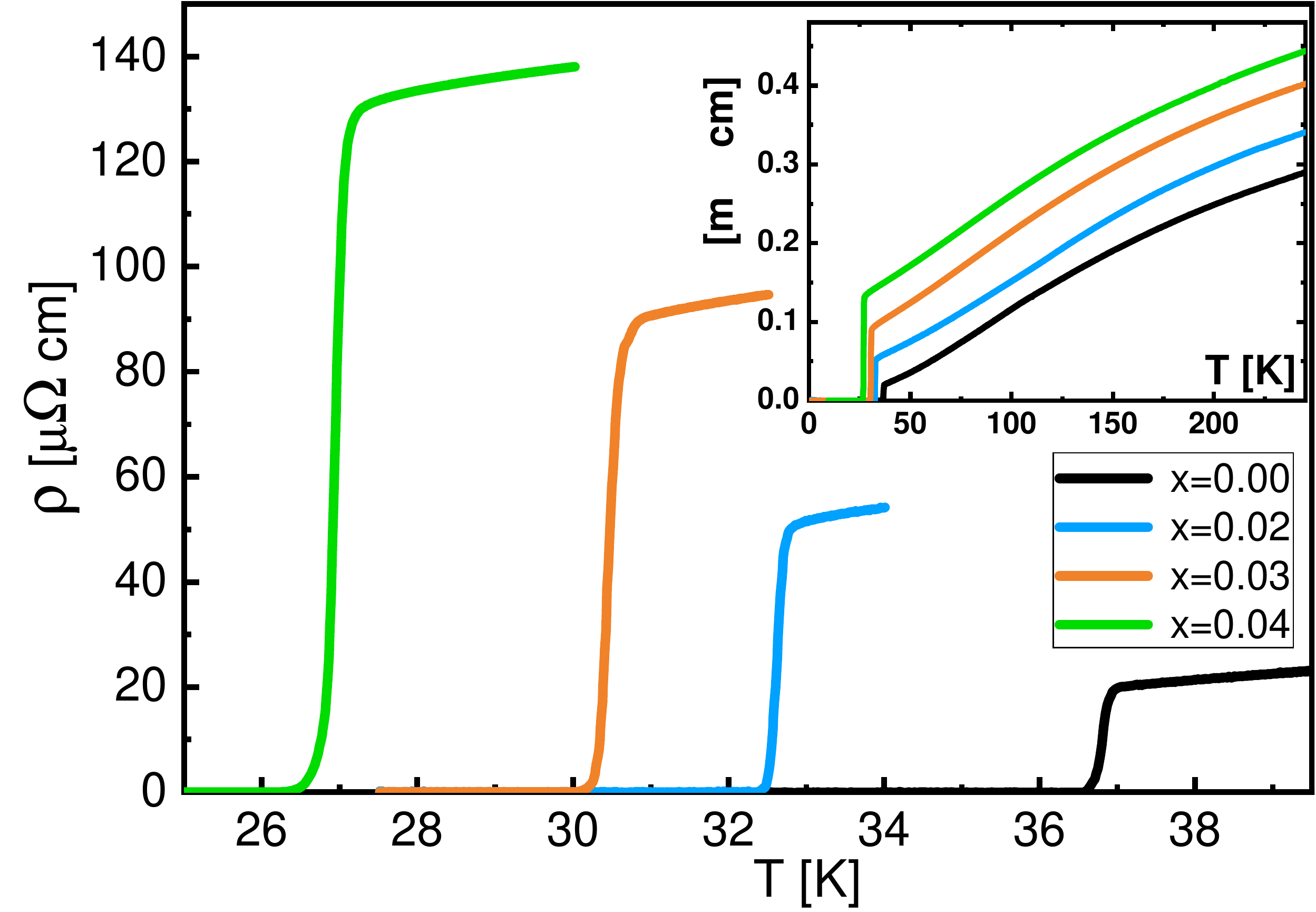}
\caption{Resistivity as a function of temperature of single crystals of {\RbEuNix} with current in-plane. As the doping increases, the normal-state resistivity increases (see inset) and the transition temperature decreases. Qualitatively the temperature dependence of the resistivity remains unchanged.
}
\label{fig:res}
\end{figure}
\section{Experiment}
\subsection{Diffraction}

We used Energy-dispersive X-ray spectroscopy (EDS) to determine the actual Ni content of the {\RbEuNix} single crystals. For this, about 10 crystals from each doping series were investigated. We found that the nominal dopings of $x = 0.03$, $x = 0.05$, and $x = 0.07$ correspond to an actual doping concentration of $x = 0.02$, $x = 0.03$, and $x = 0.04$ respectively; from here on, samples will be referred to by their average EDS doping values. The spread of doping concentrations in any batch is around $\pm$ 0.1 \%.
In XRD measurements using Cu-K$\alpha$1 radiation we observe (00l) peaks of the  of the single-crystal {\RbEuNix} with doping levels of $x = 0.00$ (black), $x = 0.02$ (blue), $x = 0.03$ (orange), $x = 0.04$ (green), as shown in Fig.\ \ref{fig:xray}. With increasing doping, the lattice peaks systematically shift to higher angles, see inset of Fig.\ \ref{fig:xray} where the (009) peak is shown. This indicates a shrinking of the $c$-axis lattice constant.
\begin{figure*}[t!]
\includegraphics[width=0.93\linewidth]{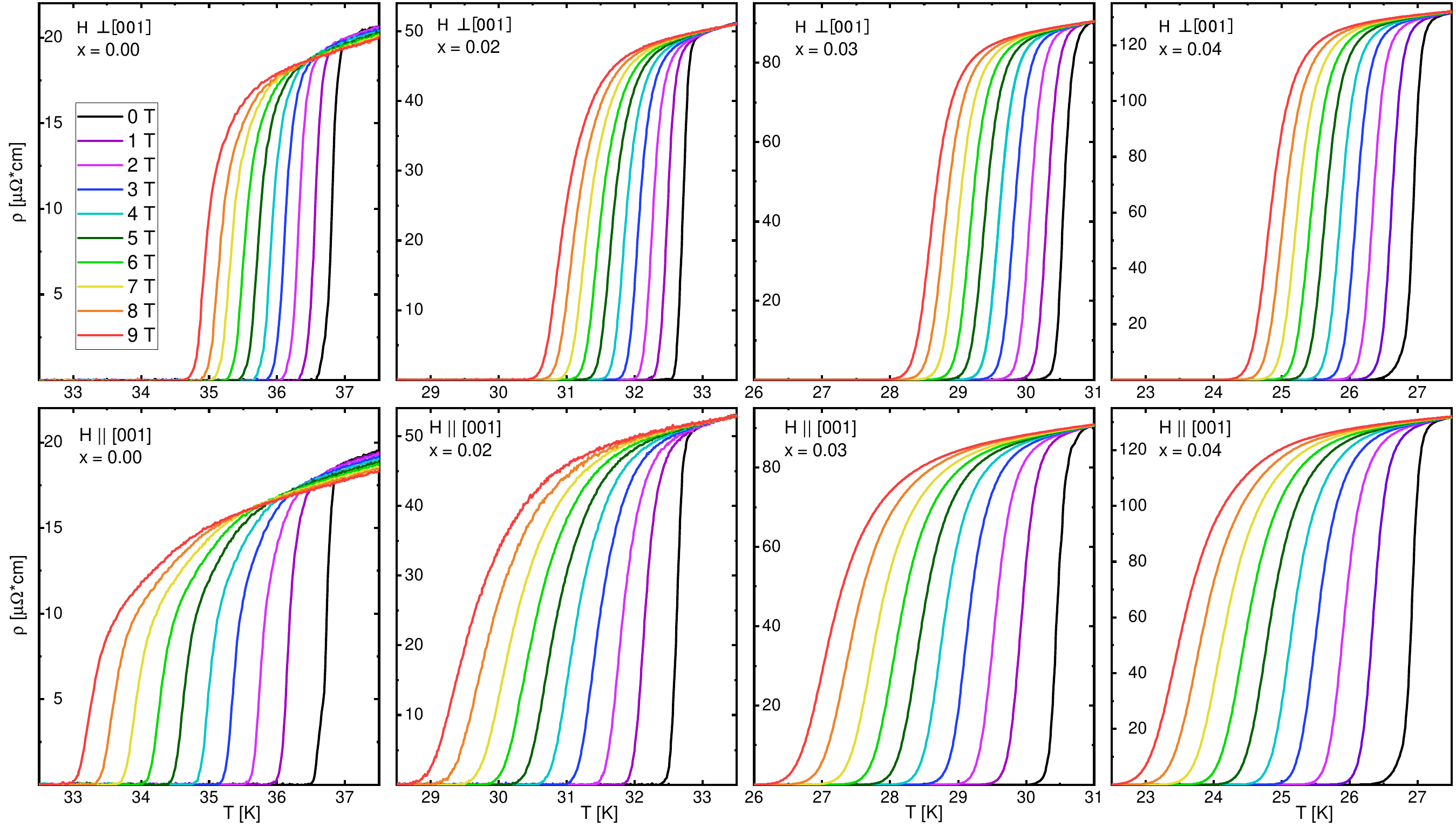}
\caption{Temperature dependence of the resistivity of {\RbEuNix} in in fields of up to 9T. Upper panels are for fields in-plane, lower panels for fields along the $c$ axis. The color code is the same for all panels. For all doping level, the superconducting transition is suppressed and broadens with increasing fields, more for fields along the $c$ axis then for in-plane fields. The suppression with fields gets larger for increasing Ni doping. The temperature axis of all panels spans 5 K whereas the vertical axis is adjusted to accommodate the large increase in resistivity with doping.
}
\label{fig:res_in_field}
\end{figure*}

We determine the $c$-axis lattice parameter from least-squares fits with a zero-shift parameter of all observed peaks as shown in Fig.\ \ref{fig:xray}.  We estimate the uncertainty to be less than $7\times10^{-4} \mathrm{\AA}$. A few samples show a small amount of nonsuperconducting Eu122-impurity phase that grows on the surface of the \RbEu crystals. No other impurity phases were observed.  The $c$ axis decreases almost linearly with increasing Ni-doping at a rate of $\approx -8\times 10^{-3} \mathrm{\AA}/\%$Ni which is in good agreement with low-doping data on polycrystalline samples \cite{Liu2017} and is consistent with Vegard's law implying the uniform incorporation of Ni.

\subsection{Resistivity}

Figure \ref{fig:res} shows the temperature dependence of the in-plane electrical resistivity $\rho_{ab}$ of single crystals of {\RbEuNix} at different doping levels. The superconducting transition temperature $T_c$ is clearly suppressed upon increasing Ni doping at a rate of $\approx$ -2.4 K/\%Ni, consistent with the polycrystalline data at low Ni-doping \cite{Liu2017}. In all samples the resistive transitions remain sharp with a transition width below 0.5 K, indicating uniform doping and single-phase material. As seen in the inset of Fig.\ \ref{fig:res}, the salient effect of Ni-doping is the remarkable parallel upwards shift of the resistivity curve. The resistivity at $T_c$ vs. $T_c$ can be seen in the inset of Fig. \ref{fig:specheat}. While the residual resistivity increases strongly, the temperature dependence of the resistivity is not altered. In particular, an upturn (on decreasing temperature) of the resistivity which was observed around 30 K on polycrystalline samples \cite{Liu2017} at doping levels of 3\% and above and interpreted as the reemergence of the spin density wave on Fe, is not apparent in our data.  As electron scattering by phonons as well as spin fluctuations depend on the details of the electronic band structure, our results imply that the effects of doping-induced changes of the band structure on electron transport are small. Recent first principals calculations \cite{Xu2019} of the electronic properties of Ni-doped {\RbEu} revealed that Ni-doping induces electron doping accompanied by an upwards shift of the Fermi energy. For 6.25\% Ni, the partial density of states derived from the Fe-d$_{x(y)z}$, d$_{z^2}$ and d$_{x^2-y^2}$ orbitals were found to decrease by approximately 10 to 20\% while for the d$_{xy}$ states it increases by 4\%.  Since $T_c$ depends exponentially on the density of states, these small changes result in a clear suppression of $T_c$.  In contrast, the resistivity depends much weaker on the density of states, and the large observed increases of the resistivity are attributable to enhanced disorder scattering due to the charged Ni-dopants. For all doping levels, we do not observe a reentrant resistive state associated with the onset of Eu sublattice magnetic order; this is in agreement with earlier reports on the polycrystalline material \cite{Liu2017} but in contrast to other Eu-containing iron-based superconductors e.g. Eu(Fe$_{1-x}$Ir$_x$)$_2$As$_2$ \cite{Paramanik2013}.

Fig.\ \ref{fig:res_in_field} shows the the superconducting transitions of the pristine and doped samples in in-plane (top row) and out-of-plane fields (bottom row). For all samples, the field-induced suppression and broadening of the superconducting transition is stronger for fields along the $c$-axis then for in-plane fields and also increases for increasing Ni content yielding a modest superconducting anisotropy of 1.8 in the pristine sample which slightly decreases upon Ni-doping (see below, Fig.\ \ref{fig:phaseboundary}). The superconducting transition temperature as a function of field was extracted as the mid-point of the transition and is used to construct the phase boundaries, which can be seen in figure \ref{fig:phaseboundary}. The high-field transitions of the pristine sample are characterized by an unusual sharpening on decreasing temperature, clearly seen in the 9T $\|$ $c$ -data.  This feature is associated with the vortex lattice melting transition \cite{Koshelev2019} and is suppressed due to the increased disorder in the doped samples. Furthermore, the negative normal state magnetoresistance that is clearly seen in the pristine sample (see also Ref. \cite{Smylie2018c}) is strongly reduced with increasing Ni-doping, consistent with the increase in impurity scattering.

\subsection{Specific Heat}
\begin{figure}[t]
\includegraphics[width=0.85\linewidth]{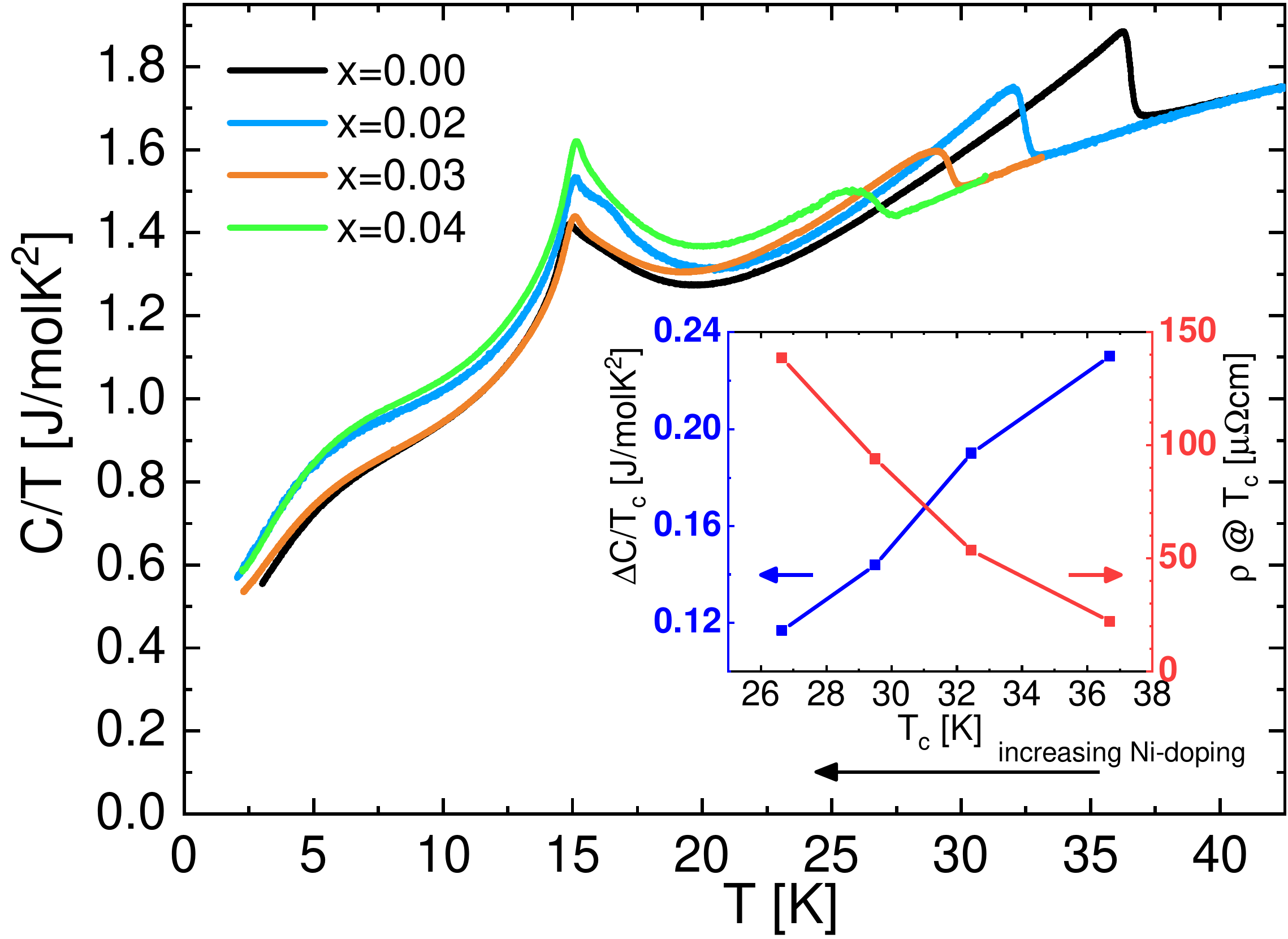}
\caption{Temperature dependence of $C/T$ of {\RbEuNix} in zero field. A clear step marks the onset of superconductivity, while the magnetic transition of the Eu moments is seen as a kink around $T_m$ = 15K.
}
\label{fig:specheat}
\end{figure}
Figure \ref{fig:specheat} shows the zero field specific heat of the pure {\RbEu} and for the three different Ni doping levels. Due to the small sample size, the determination of the molar volume for each sample introduces uncertainties. Therefore, in order to facilitate the comparison of the data from different samples, we scale the normal state molar specific heat of the Ni-doped samples to the normal state specific heat of the pure \RbEu compound. This step is motivated by the observation that substituting Fe by Ni induces essentially no change in the lattice specific heat and very little in the normal state electronic specific heat. The two clear features in $C/T$ are (i) a step signaling the superconducting phase transition at $T_c$ and (ii) a kink at the magnetic transition temperature $T_m$ while the superconducting transition temperature is clearly suppressed–both in temperature and in step-height–upon increasing Ni doping, the shape and transition temperature $T_m$ of the magnetic transition change only weakly. Even for the largest Ni concentrations investigated here, the Eu moments order below the superconducting transition. Consistent with the resistivity data, there is no additional feature in the specific heat between 2 K and room temperature that would indicate a spin-density wave (on the Fe site) as is seen in most 122 parent compounds \cite{ Johnston2010, Hirschfeld2011, Stewart2012, Tanabe2011, Hosono2015, Si2016, Bang2017, Lynn2009, Xiao2009, Koo2010}. 
\begin{figure}[t]
\includegraphics[width=0.9\linewidth]{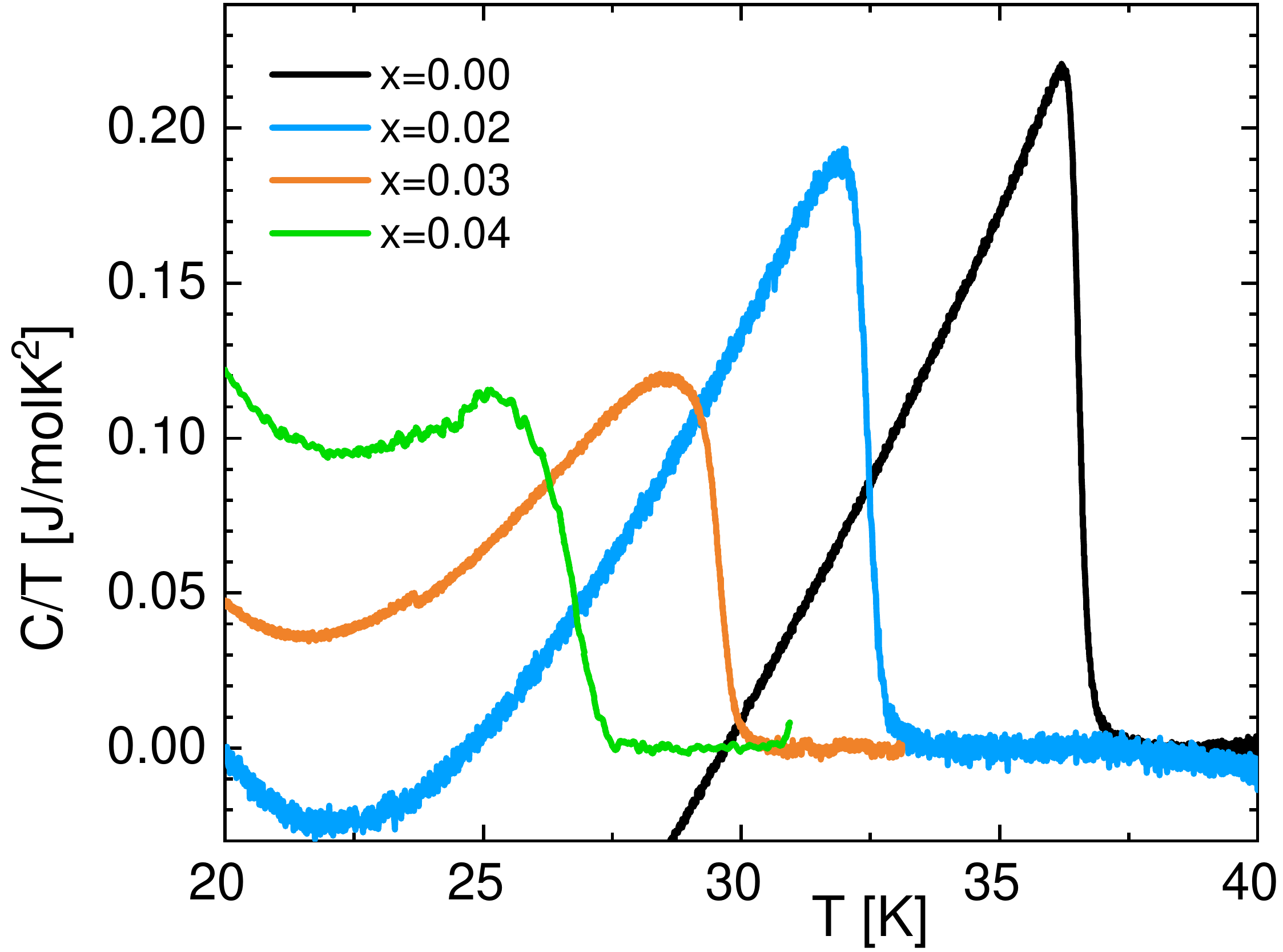}
\caption{Specific heat of the superconducting transition of {\RbEuNix} for different doping levels. A linear normal state background has been subtracted. The step height of the jump in $C/T$ decreases with increasing Ni concentration from 0.23 J/molK$^2$ down to 0.12 J/molK$^2$. The observed upturn of $C/T$ at around 23K is due to magnetic fluctuations.
}
\label{fig:specheat_Tc}
\end{figure}
\begin{figure*}[t!]
\includegraphics[width=0.93\textwidth]{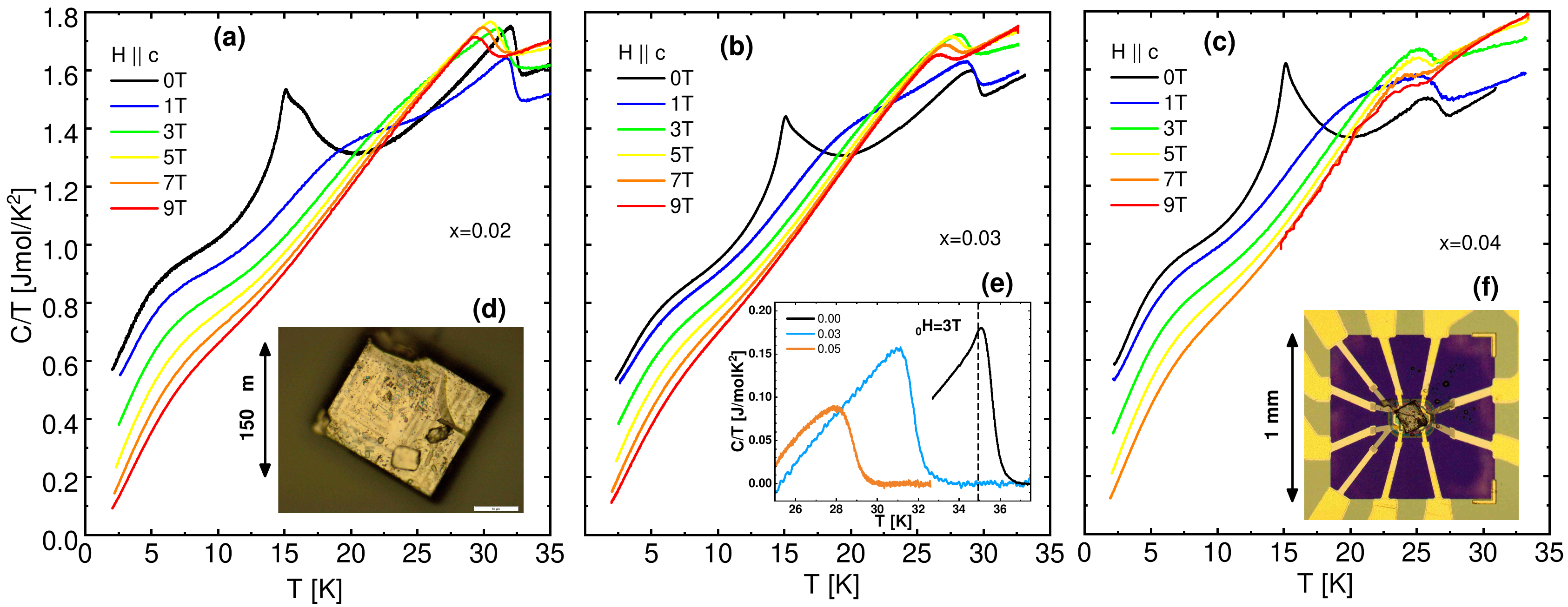}
\caption{ $C/T$ versus temperature for the three different Ni concentrations $x = 0.02$ (a), $x = 0.03$ (b) and $x = 0.04$ (c). Fields of up to 9T were applied along the $c$ axis suppressing the superconducting transition. The broadening of the transitions in field is considerably larger for higher Ni doping. The magnetic transition is transformed into a broad hump that moves to higher temperatures with fluctuations extending well above the superconducting transition. The insets (d) and (f) show a 2\% and 4\% doped {\RbEuNix} single crystal, respectively. Inset (e) shows $C/T$ (normal state background subtracted) measured in 3 T for 0, 2\% and 3\% Ni doping. In the pristine sample the additional step below $T_c$ (marked by a dotted line) indicates the vortex lattice melting. This is not visible anymore for the higher doped samples.
}
\label{fig:specheat_highH}
\end{figure*}
Figure \ref{fig:specheat_Tc} shows the superconducting contribution to the specific heat of the four samples obtained by subtracting a linear extrapolation of the normal state data above $T_c$. The transition temperature decreases from $T_c$ = 36.9K for the undoped compound to $T_c$ = 26.7K for 4\% Ni doping. While the step height is reduced from 0.23 J/mol K$^2$ to 0.12 J/mol K$^2$. The evolution of the step size with Ni-doping can be seen in inset of figure \ref{fig:specheat}. In single-band weak-coupling BCS theory, the specific heat anomaly is given as $\Delta C/T_c = 1.43 \gamma_n$, where $\gamma_n$ is the Sommerfeld coefficient of the electronic specific heat.  Thus, the rapid suppression of $\Delta C/T_c$ with Ni-doping would indicate a strong reduction in the density of states, contrary to the transport data (see above). Alternatively, however, a strong dependence of $\Delta C/T_c$ on $T_c$ has been observed in a wide variety of iron-based superconductors, a phenomenon known as the BNC scaling \cite{Budko2009}. It is believed that this behaviour arises in multi-band superconductors with strong inter-band pairing interaction \cite{Bang2016} and thus might be responsible for the strong reduction in $\Delta C/T_c$. 

Figure \ref{fig:specheat_highH} shows $C/T$ of the three Ni-doped samples in fields of up to 9T applied along the $c$-axis. In analogy to the magnetotransport data in Fig.\ \ref{fig:res_in_field}, with increasing fields, the superconducting transition is suppressed and broadened. The broadening in field is considerably larger for the higher doped samples. At the same time, the magnetic transition transforms from a kink into a crossover that moves to higher temperatures in increasing field. The associated specific heat contribution due to strong magnetic fluctuations persists far above the superconducting transition. Previous specific heat measurements on pristine \RbEu revealed a step in $C/T$ slightly below the superconducting transition \cite{Koshelev2019} that lines up with the steep drop in the resistance and is a signature of the vortex lattice melting transition. Inset (e) of Fig. \ref{fig:specheat_highH} shows $C/T$ (normal state background subtracted) measured in 3 T for 0, 2\% and 3\% Ni doping. In the pristine sample the additional step below $T_c$ (marked by the dotted line) indicates the vortex lattice melting. Upon Ni-doping, this feature is suppressed due to increased impurities and pinning sites in the sample which is consistent with the evolution of the magnetoresistance.

\begin{figure}[b!]
\includegraphics[width=0.85\linewidth]{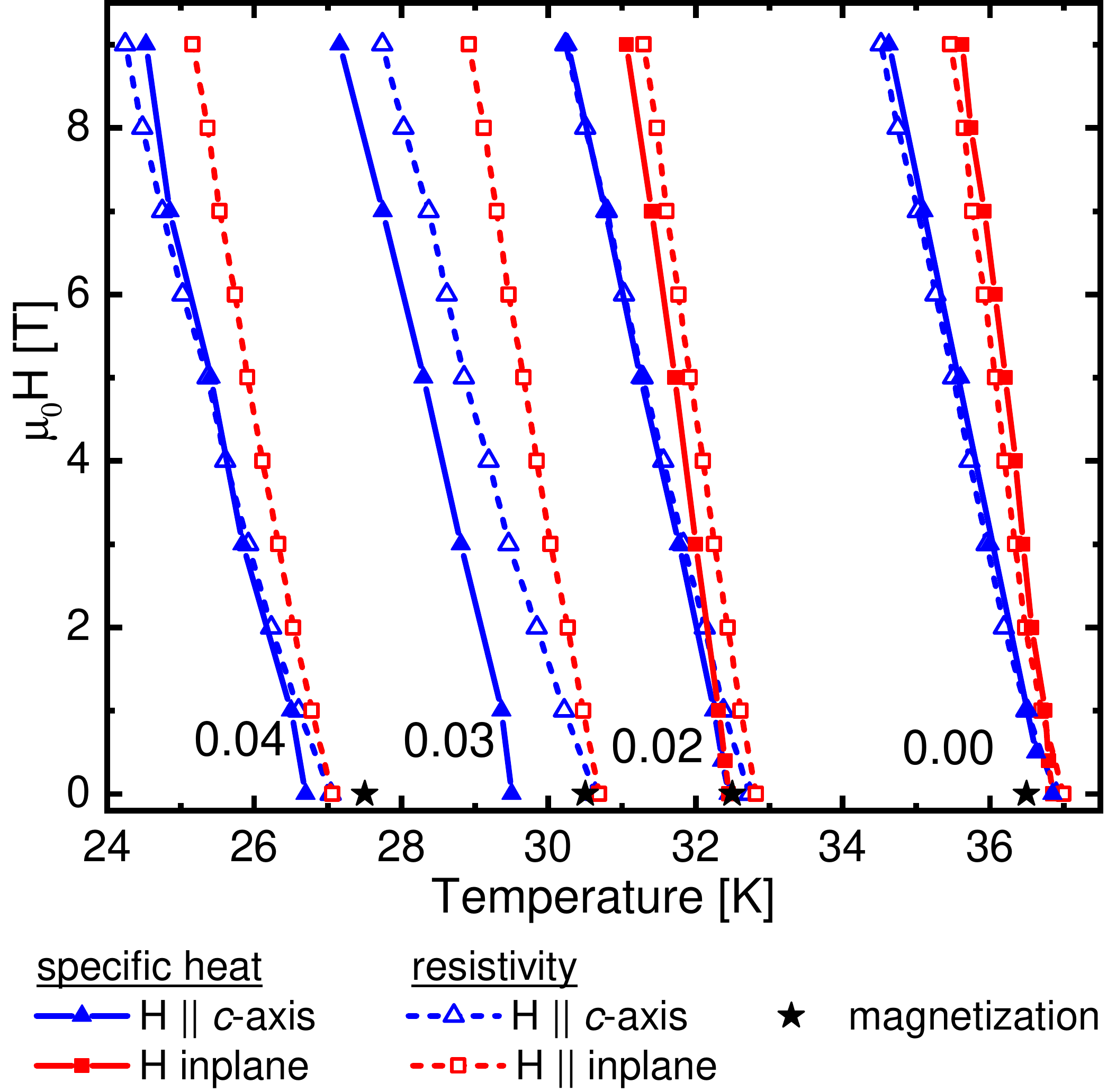}
\caption{Phase boundary of pure and Ni doped {\RbEu} single crystals for in and out-of-plane fields. Data are obtained from resistivity (open symbols and dashed lines) and specific heat (closed symbols and solid lines). The superconducting transition obtained from susceptibility measurements is indicated as well (stars).
}
\label{fig:phaseboundary}
\end{figure}
\begin{figure*}[t]
\includegraphics[width=0.93\textwidth]{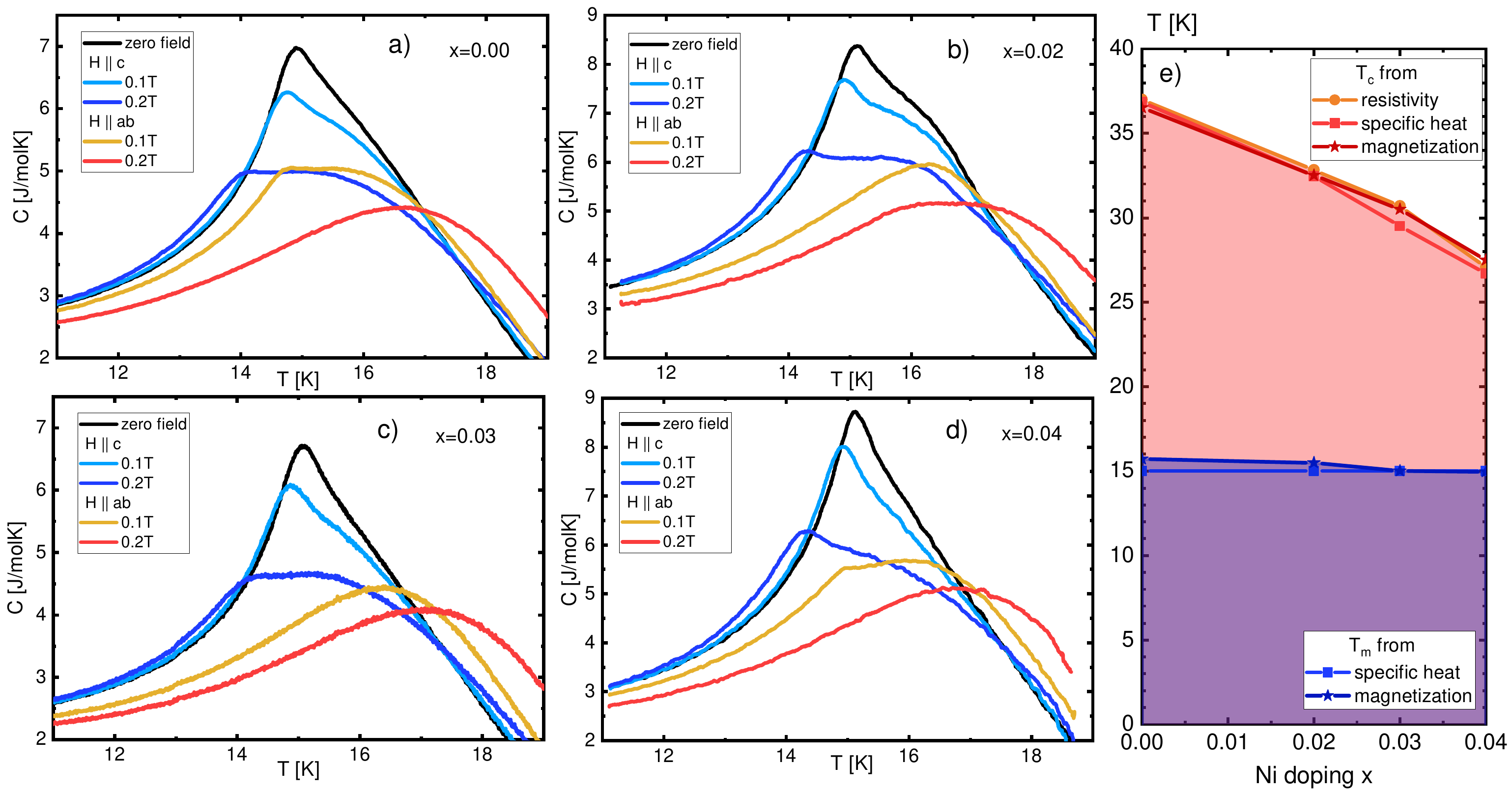}
\caption{Specific heat near the magnetic transition for 0\% (a), 2\% (b), 3\% (c), and 4\% (d) Ni doping. A 9 T background curve has been subtracted. The transition temperature $T_m = 15$ $\mathrm{K}$, here the peak in $C$, appears unchanged upon Ni doping as does the shape of the transition. Small fields along the $c$ axis shift the transition to lower temperatures while fields in-plane transform the transition into a cross-over that moves to higher temperatures. e) shows the phase diagram extracted from specific heat, magnetization and resistivity for the magnetic and superconducting transitions in zero field upon increasing Ni-doping.
}
\label{fig:specmag}
\end{figure*}
We have extracted the superconducting phase boundaries of the superconducting transition using an entropy conserving construction \cite{Willa2018} and plotted it together with those obtained from resistive mid-points and magnetization, see Fig. \ref{fig:phaseboundary}. The agreement between resistivity and specific heat derived data is generally good. The difference in $T_c$ for the 3\% sample can be attributed to measurements on crystals from different batches having slightly different doping levels. The upper critical field slopes $dH_{c2}/dT$ for both in-plane and out-of-plane fields are suppressed with increasing Ni doping. Taking the resistivity data, we observe that the $c$-axis upper critical field slope decreases from around 4 T/K for the undoped sample to 3.3 T/K for 4\% doping, whereas the in-plane direction changes from 6.3 T/K ($x = 0.00$) to $5$ $\mathrm{T/K}$ ($x = 0.04$). This results in a small reduction of the superconducting anisotropy from $\Gamma=(dH_{c2}^{ab}/dT)/(dH)_{c2}^c/dT)=1.6$ to $\Gamma=1.5$. Naively, one would expect that with increasing disorder, as is suggested by the strong increase of the residual resistivity, the upper critical fields $H_{c2}$ and $dH_{c2}/dT$ increase as the coherence length decreases. However, $T_c$ decreases also with Ni-doping. Since our data are taken close to $T_c$, we present simple estimates using the Ginzburg-Landau formalism giving $H_{c2}(T) = \Phi_0/2\pi\xi^2(T)$. In the clean limit, $\xi(0) < l = v_F\tau$ with $l$, $v_F$ and $\tau$ the electron mean free path, the Fermi velocity and scattering time, respectively, the coherence length is given by the BCS coherence length $\xi(0) = 0.74 \xi_0 = 0.76 \hbar v_F/\Delta(0) \approx \hbar v_F/2.38k_BT_c$. This results in $H_{c2}(0) \propto T_{c2}^2/v_F^2$ and an upper critical field slope $dH_{c2}/dT_c \propto T_c/v_F^2$. Thus, assuming that changes in the electronic structure are not dominant (see above), the upper critical field and the slope of it indeed decrease with decreasing $T_c$, that is, with increasing Ni-doping. The same conclusion would also be reached in the dirty limit, when the coherence length is given as $\xi(0) = 0.855 (\xi_0 l )^{1/2}$ resulting in $H_{c2}(0) \propto T_c/v_F^2\tau$. Since $\rho \propto 1/\tau$ is proportional to $T_c$ (with an offset) one again finds to leading order $H_{c2}(0) \propto T_{c2}^2$ and a slope that decreases with increasing Ni-doping.

In order to gain insight into the nature of the magnetic transition, we performed detailed specific heat measurements in the close vicinity of the transition. In zero field, the transition is characterized by a kink in the specific heat for all doping levels, a non-singular behavior that was already observed for the undoped compound and identified as a BKT transition \cite{Willa2019}. As noted above, the transition temperature of $T_m = 15$ $\mathrm{K}$ is largely unaffected by the Ni doping. Applying small fields along the $c$ axis shifts the magnetic transition to lower temperatures while applying the field in the plane replaces the transition with a crossover. The very similar field-response of the specific heat of the doped samples (Fig. \ref{fig:specmag}) and of the undoped samples (Ref. \cite{Willa2019}), leads us to conclude that Ni-doped {\RbEu}, just as the stoichiometric compound,  belongs to the universality class of the anisotropic 2D XY model, with the Eu-moments having an easy-plane anisotropy, strong ferromagnetic intralayer coupling and very weak interlayer interactions leading to the BKT transition of the moments in the plane near $T_m = 15$ K. This transition will be suppressed to lower temperatures for fields along the $c$ axis, whereas it will be replaced with a broad crossover at higher temperatures for fields in the plane since any field in the plane destroys the rotational symmetry necessary for a BKT transition. 

%

\subsection{Magnetization}
\begin{figure}[t]
\includegraphics[width=0.85\linewidth]{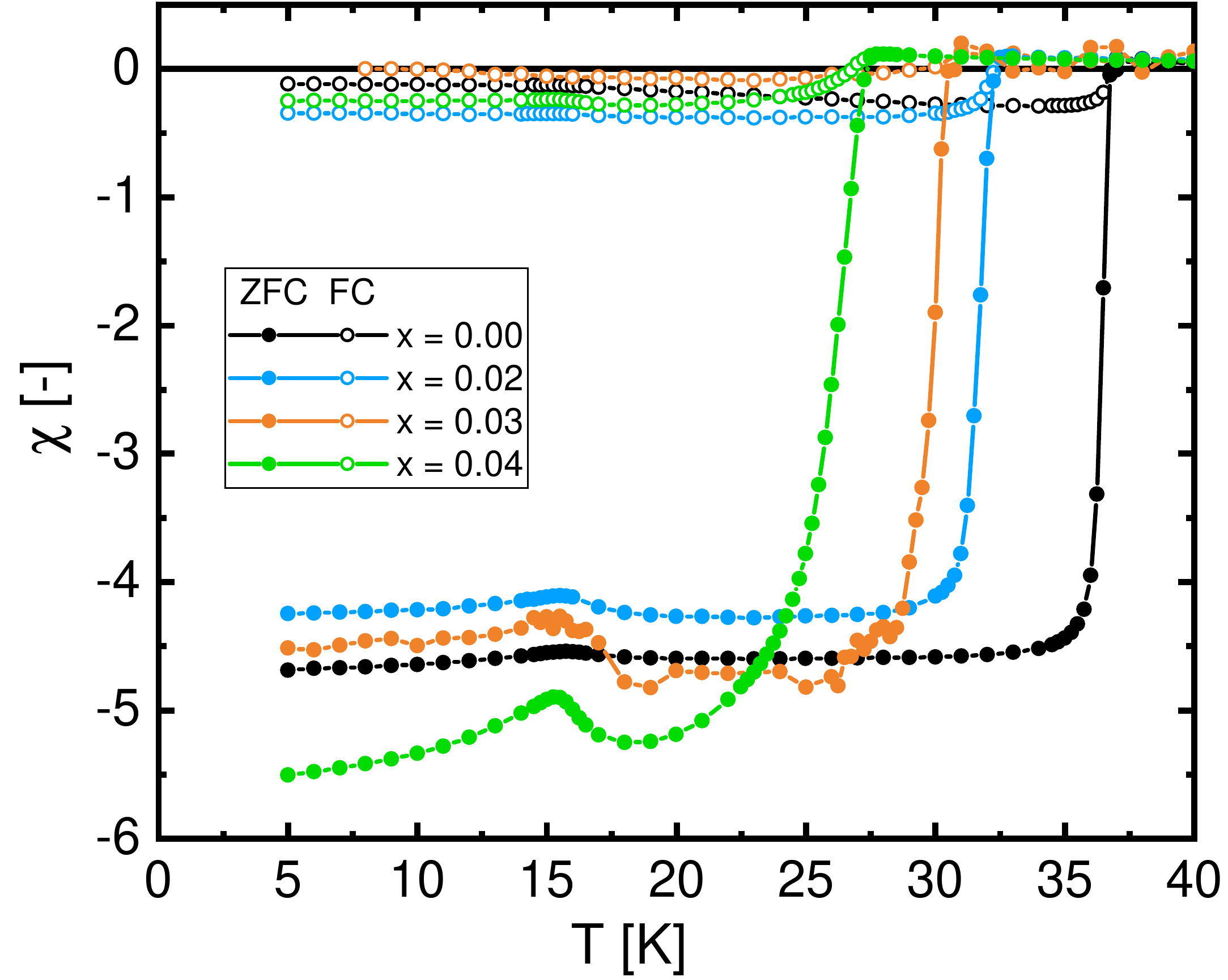}
\caption{The ZFC and FC magnetic susceptibility of {\RbEuNix} single crystals with $H = 1 $mT along the [001] direction. With increasing doping, the onset $T_c$ shifts lower while the transitions remain parallel, indicating homogenous doping. In this orientation, the diamagnetism masks most of the Eu magnetism.}
\label{fig:Tc_mag}
\end{figure}

We study the magnetic state of {\RbEuNix} by measuring the zero field-cooled (ZFC) and field cooled (FC) temperature dependence of the magnetization in different applied fields as well as the field dependence of the magnetization at different temperatures with the field applied perpendicular or parallel to the $c$ axis. Fig.\ \ref{fig:Tc_mag} shows ZFC and FC measurements of magnetic susceptibility, $\chi = M / H$, for $x = 0.00$, $0.02$, $0.03$, and $0.04$ with $H = 10 \mathrm{G}$ applied along [001]. Slight differences in demagnetization factors may cause variations in the absolute magnitude of susceptibility. As doping increases, the onset of diamagnetism---taken as $T_c$---monotonically decreases from $36.5 \mathrm{K}$ to $27.5 \mathrm{K}$, in agreement with the resistivity and the specific heat data. Due to the plate like sample geometries, magnetization data for $H \| [001]$ contains large superconducting contributions, making the Eu magnetism difficult to discern in this geometry.

\begin{figure}[H]
\includegraphics[width=0.93\textwidth]{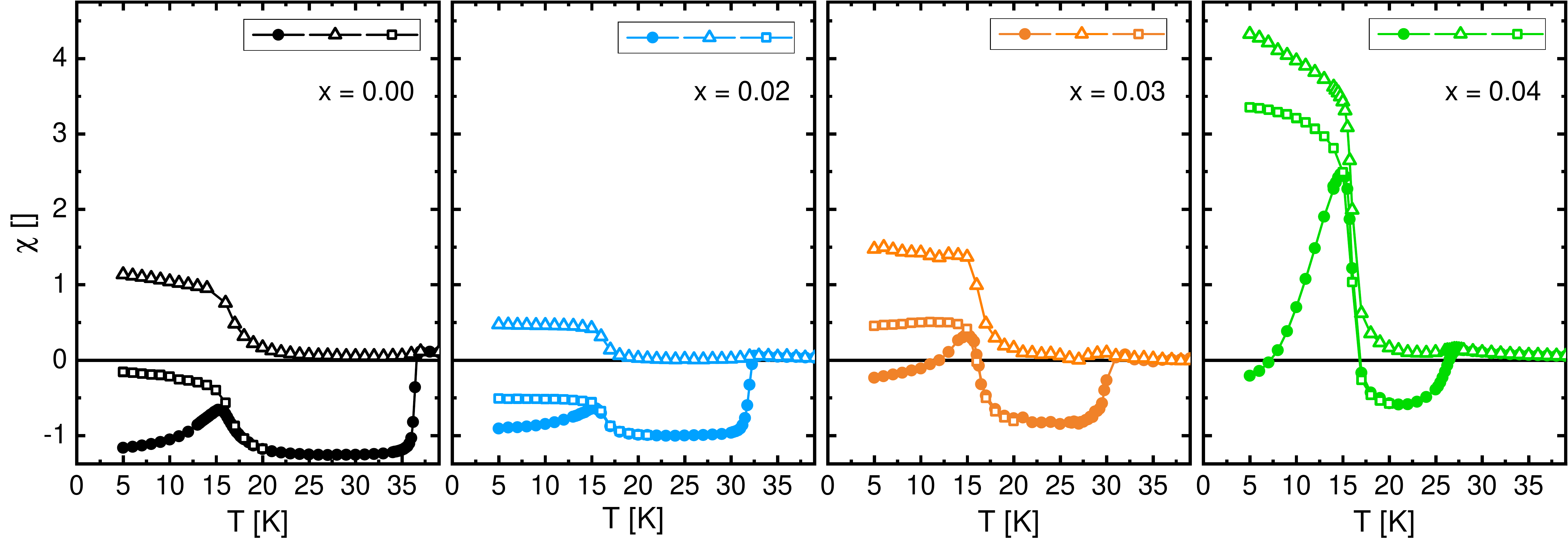}
\caption{The ZFC and FC magnetic susceptibility of {\RbEuNix} single crystals with $H = 1$ $\mathrm{mT}$ applied along the [110] direction. Solid circles mark ZFC, hollow triangles, FC. Hollow squares represent cooling in zero field, applying $10$ $\mathrm{mT}$, warming to $20$ $\mathrm{K}$, and then measuring FC. As doping increases, the peak of the magnetic transition is almost unchanged showing that magnetism is essentially unaffected by Ni doping.
}
\label{fig:Mag_mag}
\end{figure}

Applying magnetic fields along [110] decreases the contribution of diamagnetism, allowing for a more pronounced development of the Eu magnetism, as shown in Fig.\ \ref{fig:Mag_mag}. In the magnetization measurements, we find that $T_m$, defined as the peak in the ZFC susceptibility below $T_c$, is slightly suppressed from $15.3 \mathrm{K}$  to $15.0 \mathrm{K}$ by low Ni doping. This is in rough agreement with the specific heat data showing that the magnetic ordering temperature is essentially independent of Ni-doping. 

\begin{figure}[H]
\centering
\includegraphics[width=0.9\textwidth]{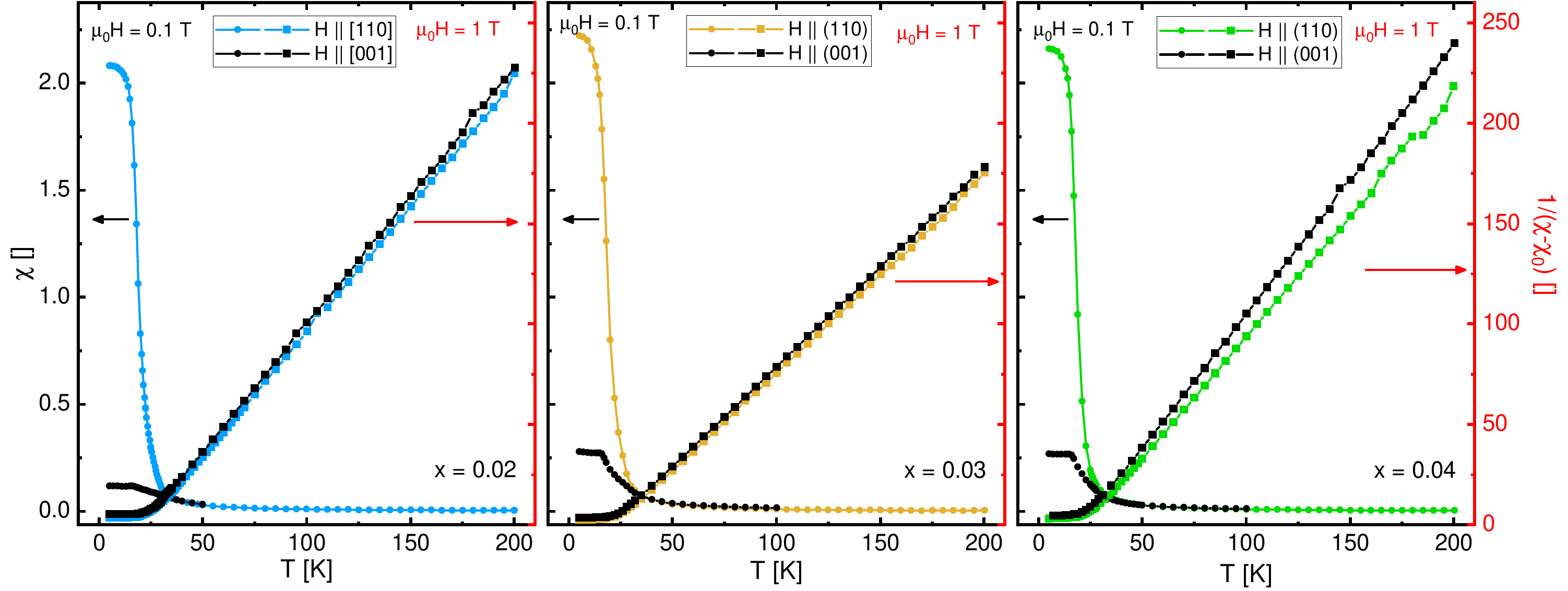}
\caption{Temperature dependence of field-cooled magnetic susceptibility measured upon warming in {\RbEuNix} single crystals: (a) $x = 0.02$. (b) $x = 0.03$. (c) $x = 0.04$. For all panels, colored symbols represent $H \| [110]$, black symbols represent H $\|$ [001]. For $\mu_0 H = 0.1 \mathrm{T}$, the large in- to out-of-plane anisotropy is present at all doping levels. The right-hand axis in each panel plots the inverse susceptibility measured for fields of 10 T; the low-temperature behavior is more isotropic, and the high-temperature ($>50 \mathrm{K}$) data can be well described by Curie-Weiss behavior.
}
\label{fig:CurieWeiss}
\end{figure}

In higher fields, the magnetic susceptibility for FC measurements at all doping levels is qualitatively very similar to the undoped compound \cite{Smylie2018c} as shown in Fig.\ \ref{fig:CurieWeiss}. In 0.1 T we observe a large anisotropy in the low temperature susceptibility with $\chi_{ab} \gg \chi_{c}$ under FC conditions for which vortex pinning effects are small. Thus, the strong easy-plane anisotropy of the Eu-moments observed in the undoped compound is preserved with Ni-doping. At 1 T and above $50$ $\mathrm{K}$ the data are well described by a Curie-Weiss law $\chi(T) = \chi_0 + C/(T-\Theta_c)$, yielding $\Theta_c$ near $23$ $\mathrm{K}$ and an effective moment $\mu_{\mathrm{eff}} = 2.837 C^{1/2}$ close to the expected Eu$^{2+}$ effective moment of $\mu_{\mathrm{eff}} = g \mu_B \sqrt{S(S+1)} = 7.94 \mathrm{\mu_B /Eu}$ (with g = 2 and S = 7/2) suggesting that these low levels of Ni doping do not substantially affect the Eu magnetic interactions as expected in numerical simulations \cite{Xu2019}. Fit values are shown in table \ref{tab:data}. The higher value of $\mu_{\mathrm{eff}}$ found for $x = 0.03$ is likely due to surface EuFe$_2$As$_2$ phase on the sample, which would contribute twice as much Eu per volume as the 1144 phase would. The positive Curie-Weiss temperature values suggests predominantly ferromagnetic interactions between the Eu-moments consistent with what has been seen in the undoped compound. Within a 2D Heisenberg model that includes in-plane versus out-of-plane spin anisotropy \cite{Willa2019}, the anisotropy of the Curie-Weiss temperatures between in and out-of-plane fields is a measure of the anisotropy of the Eu spins. It is essentially unchanged upon Ni doping consistent with the data obtained at low temperatures.

\onecolumngrid

\vfill
\vfill
\pagebreak

\vfill
\begin{figure}[H]
\centering
\includegraphics[width=0.9\textwidth]{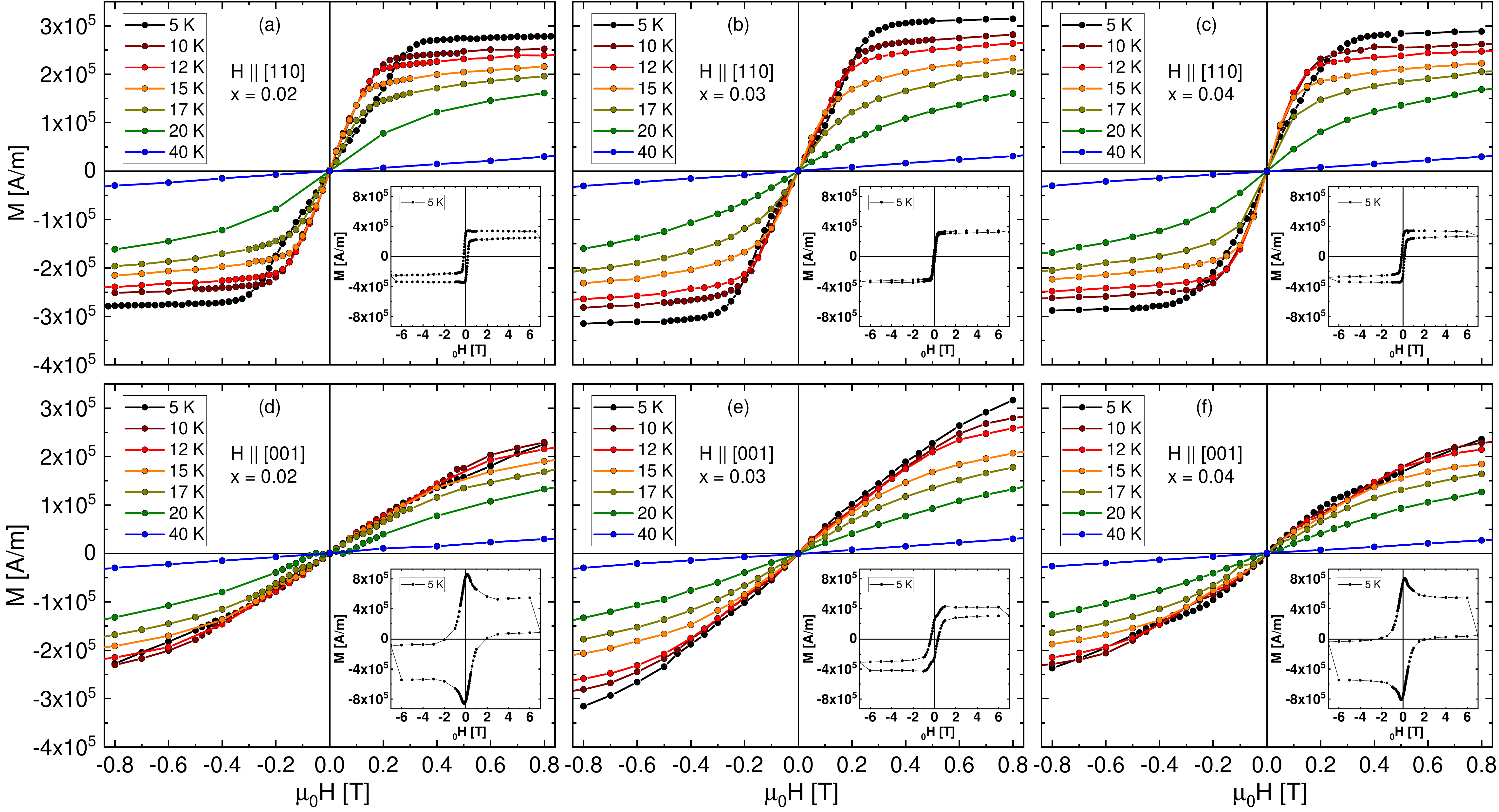}
\caption{Magnetization of the Eu-sublattice vs applied field H $\|$ [110] (a-c) and along H $\|$ [001] (d-f) for different doping levels and at various temperatures. The insets show the as-measured magnetization hysteresis loops. At high fields, the saturation magnetization approaches $\approx$ 300 emu/cm$^3$, approximately that of the pristine material. In all measurements, the slope of the magnetization increases as $T_m$ is approached from above, and decreases for $T < T_m$ as the magnetic lattice becomes stiffer.
}
\label{fig:hysteresis}
\end{figure}
\pagebreak

\begin{table}[H]
\centering
\begin{tabular}{|c|c|c|c|c|c|c|c|c|c|}
\hline 
	\makebox[3em]{$x$}
	& \makebox[4.5em]{$c$}
	& \makebox[4.5em]{$T_C$}
	& \makebox[4.5em]{$\rho$ @ $T_{c}$}
	& \makebox[4.5em]{$\Delta C/T_c$}
	& \makebox[4.5em]{$\mu_{\mathrm{eff}}$}
	& \makebox[4.5em]{$\Theta_C^{xy}$}
	& \makebox[4.5em]{$\Theta_C^z$}
	& \makebox[4.5em]{$\Theta_C^{xy}/\Theta_C^z$}
	& \makebox[4.5em]{$M_{sat}$}
	\tabularnewline
	$[-]$&[\AA] & [K] &
	[$\mu\Omega$ cm] & [J/molK$^2$] & [$\mu_B$/Eu] & [K] & [K] & $[-]$ & [$\mu_B$/Eu]
	\tabularnewline
	\hline 
	\hline 
	0.00 & 13.3047 & 36.9 & 22 & 0.23 & 7.75 & 23.00 & 21.40 & 1.075 & 7.2\tabularnewline
	\hline 
	0.02 & 13.2911 & 32.5 & 54 & 0.19 & 7.87 & 23.18 & 21.94 & 1.057 & 6.4\tabularnewline
	\hline 
	0.03 & 13.2825 & 30.7 & 94 & 0.14 & 8.89 & 22.53 & 20.87 & 1.078 & 8.1\tabularnewline
	\hline 
	0.04 & 13.2733 & 26.9 & 139 & 0.12 & 7.97 & 22.84 & 21.04 & 1.086 & 6.5\tabularnewline
	\hline 
\end{tabular}
\caption {Properties of single crystals of {\RbEuNix} derived from x-ray diffraction, resistivity, specific heat and magnetization measurements. $\Theta_C^{xy}$ denotes the Curie temperature for in-plane fields and $\Theta_C^z$ the one for fields along the $c$ axis.
\label{tab:data}}
\end{table} 
\twocolumngrid

\pagebreak
\begin{figure}[b]
\includegraphics[width=0.9\linewidth]{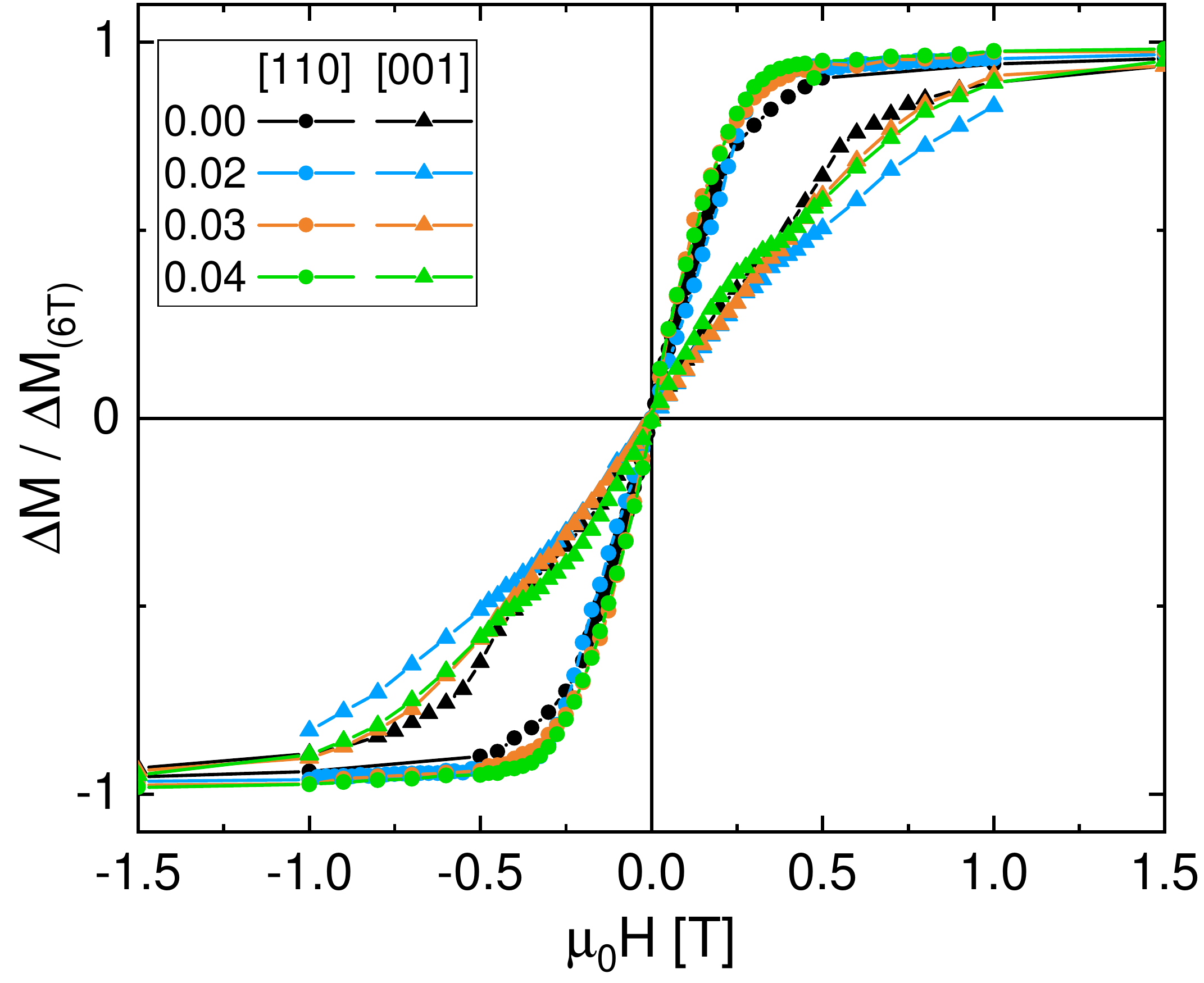}
\caption{Comparison of the symmetrized component, representing the ferromagnetic component, of the magnetic hysteresis of Ni-doped single crystals of {\RbEuNix} with H $\|$ [110] (circles) and H $\|$ [001] (triangles). Increasing the doping does not change the field necessary to collapse the Eu magnetic lattice. \vspace{-1.8em}
}
\label{fig:symmetrized}
\end{figure}

The insets of Figs. \ref{fig:hysteresis} (a)-(f) show magnetization hysteresis loops measured at 5 K for in and out-of-plane fields. The superposition of a hysteretic superconducting signal and a ferromagnetic-like signal is clearly seen, especially for H $\|$ [001]. The large sample cross-section and high critical current density for H $\|$ [001] makes this an expected behavior. In comparison to the $x = 0.02$ and $x = 0.04$ samples, the $x = 0.03$ sample displays a lower superconducting hysteresis corresponding to weak pinning which suggests less disorder.  Assuming that the superconducting hysteresis is symmetric about the equilibrium magnetization curve and that the magnetic hysteresis of the Eu spins is negligible as was found in the nonsuperconducting parent compound EuFe$_{2}$As$_{2}$ \cite{Jiang2009} we determine the magnetization curve M(H) of the Eu spin sublattice as $M = (M_+ + M_-)/2$ where $M_+ (M_-)$ is the magnetization measured in increasing (decreasing) applied field. The results are shown for multiple temperatures in the main panels of Figs. \ref{fig:hysteresis} (a)-(f). Above $T_m$, a Brillouin-like response is observed. The slopes of the magnetization curves grow as $T_m$ is approached, then become slightly lower as $T$ decreases and the magnetic lattice stiffens. For H $\|$ [110] and H $\|$ [001], the saturation magnetization reaches $\approx$ 300 emu/cm$^3$ (compare table \ref{tab:data}) corresponding to 6.4 $\mu_B$/Eu, comparable to the expected full moment. Some excess moment was found for the $x = 0.03$ sample, that can again be attributed to an impurity EuFe$_2$As$_2$ phase. Additionally, differences from the free moment value may arise from the approximation of the hysteresis model.

The 5 K Eu-sublattice magnetization curves for the three doping levels and a pristine sample in both field orientations are plotted together in Fig.\ \ref{fig:symmetrized}. In order to minimize the effect of variations from varying degrees of hysteresis and from uncertainties in sample volume, we normalize each curve to its value at 6 T. There is no systematic change with doping of the saturation field in either orientation, with saturation fields of $H_{\mathrm{sat}}^{ab} \approx 0.21 \mathrm{T}$ and $H_{\mathrm{sat}}^{c} \approx  0.8 $-$ 1 \mathrm{T}$ for all doping levels, consistent with the pronounced easy-plane magnetic anisotropy.\vspace{-1.4em}

\section{Conclusion}

We have studied the effect of up to 4\% Ni doping on the Fe-site in the magnetic superconductor {\RbEu} in specific heat, resistivity and magnetization measurements. We observe a clear suppression of the superconducting transition temperature $T_c$. Upon Ni-doping, the resistivity curve shifts up in a parallel fashion indicating a strong increase of the residual resistivity due to scattering by charged dopant atoms while the electronic structure appears largely unchanged. The observed step $\Delta C/T_c$ at the superconducting transition decreases strongly for increasing Ni doping in agreement with BNC scaling. The upper critical field slopes are reduced upon Ni doping for in- as well as out-of-plane fields leading to a small reduction in the superconducting anisotropy. Specific heat measurements of the magnetic transition reveal the same BKT behavior close to the transition temperature $T_m$ for all doping levels with the transition temperature essentially unchanged and the in- to out-of-plane anisotropy of Eu-magnetism observed at small magnetic fields is unaffected. All of these observations indicate a decoupling of the Eu magnetism from superconductivity and essentially no influence of Ni doping on the Eu magnetism in this compound. These findings are in line with recent first principles calculations \cite{Xu2019} indicating that the coupling between the Eu-moments is mediated by the RKKY interaction via the Fe-d$_{z^2}$ orbitals which were found to be remarkably insensitive to Ni-doping.

\section{Acknowledgment}

This work was supported by the U.S.\ Department of Energy, Office of Science, Basic Energy Sciences, Materials Sciences and Engineering Division. K.\ W. acknowledges support from the Swiss National Science Foundation through the Early Postdoc.Mobility program.

\end{document}